\documentclass[10pt,a4paper,english]{extarticle}
\usepackage[T1]{fontenc}
\usepackage[latin9]{inputenc}
\usepackage{babel}
\usepackage{graphicx}
\usepackage{esint}
\usepackage[unicode=true]
 {hyperref}
\usepackage{breakurl}

\makeatletter

\usepackage{meta}
\usepackage{siunitx}
\usepackage{epsfig}
\hypersetup{hidelinks}

\title{Transfer-matrix method for second-order nonlinear processes with realistic beams}

\makeatletter
\def\name#1{\gdef\@name{#1\\}}
\makeatother
\name{{\bf \large A. Loot* and V. Hizhnyakov}}

\address{
Institute of Physics, \\
University of Tartu, Tartu, Estonia \\
*corresponding author, E-mail: {\tt ardi.loot@ut.ee}
}

\let\oldbibliography\thebibliography
\renewcommand{\thebibliography}[1]{%
  \oldbibliography{#1}%
  \setlength{\itemsep}{0pt}%
}

\makeatother

\begin{document}
\maketitle
\begin{abstract}
Accurate and fast modeling of electric fields in layered structures
have a great scientific and practical value. Prevalent method for
that is transfer-matrix method. However, transfer matrix method is
limited to infinite plane wave calculations, which can become a limiting
factor if a very narrow resonances, e.g. long range surface plasmon
polaritons, are present in the structure. In this paper we extend
the functionality of standard and nonlinear transfer-matrix method
to include beams with arbitrary profile and propose applications for
the method. 
\end{abstract}

\section{Introduction}

Standard transfer-matrix method (TMM) is a very powerful and a fast
method to solve the Maxwell equations in a layered structures \cite{Abeles1957,Chilwell1984}.
However, standard TMM is limited to linear optics and calculation
of plane waves. In our previous paper (Ref. \cite{Loot2017}) we extended
standard TMM to second-order nonlinear processes (NLTMM) for calculation
of plasmonic structures. This paper is devoted to extending the both
standard and nonlinear TMM to use realistic beams with arbitrary profile.
This communication will include all the details of the formulation
used to extend the method. The improved method itself is made publicly
available at \href{https://github.com/ardiloot/NonlinearTMM}{github.com/ardiloot/NonlinearTMM}. 

The paper will begin with a short review to a standard TMM (Sec.\ref{subsec:tmm-Review}).
Next the functionality of TMM is first extended to calculate the fields
of the beam with arbitrary profile (Sec. \ref{subsec:TMM for waves-fields}
and \ref{subsec:Fields-in-layered}) and then the theory for efficient
calculation of the powers of the beams is developed (Sec. \ref{subsec:tmm-for-waves-pwr-s}
and \ref{subsec:tmm for waves-pwr-p}). Next, similar functionality
is added to nonlinear TMM (Sec. \ref{subsec:NLTMM}). Finally, the
paper is concluded and the possible applications of the method is
outlined.

\section{Standard TMM\label{sec:Standard-TMM}}

This section will focus on the extending the standard TMM to employ
realistic beams with arbitrary profile. This section begins with a
short review to TMM (for more details see Refs. \cite{Abeles1957,Chilwell1984,Loot2017}),
in Sec. \ref{subsec:TMM for waves} the theory to use realistic beams
will be developed. The theory developed for standard TMM will be used
in Sec. \ref{subsec:NLTMM} to extend the NLTMM.

\subsection{Review\label{subsec:tmm-Review}}

\begin{figure}[h]
\centering{}\includegraphics[width=5cm]{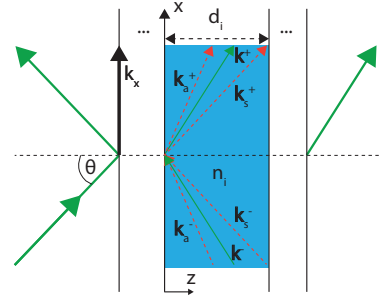}\caption{The usual layered structure for TMM calculations. The layers are defined
by the thicknesses $d_{i}$ and refractive indices $n_{i}.$ The input
beam is incident from the left at a angle of incidence $\theta$ which
correspond to tangential wave vector $k_{x}.$ In the case of standard
TMM we have two plane waves $\mathbf{k}^{\pm}$ is a single layer,
however in case of NLTMM we have in addition nonlinear inhomogeneous
waves denoted by $\mathbf{k}_{a}^{\pm}$ and $\mathbf{k}_{s}^{\pm}.$\label{fig:nltmm-structure}}
\end{figure}

The usual layered structure of TMM simulation is shown in Fig. \ref{fig:nltmm-structure}.
It consists of $N+1$ layers of different thickness $d_{i}$ and refractive
index $n_{i}.$ In every layer the electrical fields are described
as a sum of forward- and backward-propagating waves (layer index $i$
omitted)

\begin{equation}
\mathbf{E}=\mathbf{E}^{+}+\mathbf{E}^{-}=\mathbf{A}^{+}e^{i\mathbf{k}^{+}\cdot\mathbf{r}}+\mathbf{A}^{-}e^{i\mathbf{k}^{-}\cdot\mathbf{r}},\label{eq:frw-bkw}
\end{equation}

\noindent where $+$ and $-$ denote the forward- and backward-propagating
waves, $\mathbf{A}^{\pm}$ is the amplitude, $\mathbf{k}^{\pm}$ is
the wave vector and $\mathbf{r}$ is a standard position vector. The
change of the amplitudes $\mathbf{A}^{\pm}$ of the plane waves inside
a single layer is described by a propagation matrix and the continuity
relations on the boundary of the different layers are forced by transfer
matrix \cite{Loot2017}.

The input beam is incident from the left (see Fig. \ref{fig:nltmm-structure})
under angle of incidence $\theta,$ which corresponds to the tangential
wave vector component $k_{x}=2\pi n_{0}/\lambda$ (same in every layer),
where $\lambda$ is the wavelength of the light in vacuum.

\subsection{Realistic beams\label{subsec:TMM for waves}}

Standard TMM works with infinite plane waves, here we present the
formulation to use beams with arbitrary profile. First, we develop
the theory to calculate the field distribution in the case of realistic
beams in a infinite homogeneous medium and then move to the calculation
of the fields in the stratified structures. An extra care must be
devoted to calculate the powers of arbitrary beams and it is explored
in Sec. \ref{subsec:tmm-for-waves-pwr-s} and \ref{subsec:tmm for waves-pwr-p}.

\subsubsection{Fields in a infinite medium\label{subsec:TMM for waves-fields}}

\begin{figure}[h]
\centering{}\includegraphics[width=5cm]{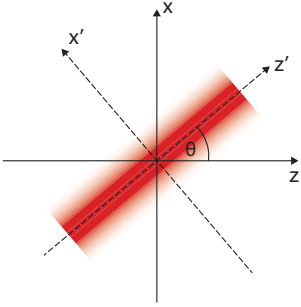}\caption{The coordinate system of the wave ($x',z'$) and the coordinate system
of the structure ($x,z$). The wave is propagating at the angle of
incidence $\theta$ in the structure. \label{fig:waves-coordinates}}
\end{figure}

Lets look a wave with an arbitrary cross-sectional electrical field
(wave profile) $\mathbf{E}_{w}\left(x'\right)=\mathbf{E}\left(x',z'=0\right)$
(defined at $z'=0$) propagating in infinite homogeneous a medium
under angle of incidence $\theta$ (see Fig. \ref{fig:waves-coordinates}).
Then, in the coordinate system of the wave ($x'$, $z'$) we can easily
express the electrical field through angular spectrum representation
(see Ref. \cite{Novotny2006}) as

\begin{equation}
\mathbf{E}\left(x',z'\right)=\int_{-k}^{k}dk_{x}'\,\mathbf{\check{E}}_{w}\left(k_{x}'\right)e^{i\left(k_{x}'x'+k_{z}'z'\right)},
\end{equation}

\noindent where $\left(k_{x}',k_{z}'=\sqrt{k^{2}-k_{x}'^{2}}\right)$
are the wave vector x and z components in the wave coordinate system,
$k$ is the wave number and $\mathbf{\check{E}}_{w}\left(k_{x}'\right)$
is the Fourier transform of the wave profile

\begin{eqnarray}
\mathbf{\check{E}}_{w}\left(k_{x}'\right) & = & \mathcal{F}\left[\mathbf{E}_{w}\left(x'\right)\right]=\nonumber \\
 & = & \frac{1}{2\pi}\int_{-\infty}^{\infty}dx'\,\mathbf{E}_{w}\left(x'\right)e^{-ik_{x}'x'}.\label{eq:waves-fft}
\end{eqnarray}

\noindent After rotation of coordinates by $-\theta$ we get

\begin{equation}
\mathbf{E}\left(x,z\right)=\int_{-k}^{k}\frac{dk_{x}}{\cos\left(\theta\right)}\mathbf{\check{E}}_{w}\left(k_{x}'\right)e^{i\left(k_{x}x+k_{z}z\right)},\label{eq:tmm-pw-exp}
\end{equation}

\noindent where $k_{x}'=\cos\left(\theta\right)k_{x}-\sin\left(\theta\right)k_{z}.$
Eq. \ref{eq:tmm-pw-exp} together with Eq. \ref{eq:tmm-pw-exp} has
a significant meaning: if we know the beam cross-sectional profile
$\mathbf{E}_{w}\left(x'\right)$ we can find the electrical field
of the wave everywhere under any angle of incidence $\theta$ by integration
of plane waves in Eq. \ref{eq:tmm-pw-exp}. This result must be now
generalized to layered structures.

\subsubsection{Fields in layered medium\label{subsec:Fields-in-layered}}

Noticing, that $\mathbf{\check{E}}_{w}\left(k_{x}'\right)e^{i\left(k_{x}x+k_{z}z\right)}$
is just a plane wave with amplitude of $\mathbf{\check{E}}_{w}\left(k_{x}'\right),$
which is easily calculated by standard TMM, it is evident, that using
Eq. (\ref{eq:tmm-pw-exp}) allows to extend TMM for waves with arbitrary
profile $\mathbf{E}_{w}\left(x'\right)$ through a single integral.
In other words, for the input wave the amplitudes $\mathbf{\check{E}}_{w}\left(k_{x}'\right)$
are know from Eq. \ref{eq:waves-fft}, and for the output waves the
corresponding amplitudes are easily calculated by TMM and could be
converted to final fields by Eq. \ref{eq:tmm-pw-exp} by integration.

In the case of Gaussian input wave the Fourier transform of is beam
profile is given analytically by

\begin{equation}
\mathbf{\check{E}}_{w}\left(k_{x}\right)=\mathbf{E}_{0}\frac{w_{0}}{2\sqrt{\pi}}e^{-k_{x}^{2}w_{0}^{2}/4},
\end{equation}

\noindent where $\mathbf{E}_{0}$ is the amplitude of the wave and
$w_{0}$ is the waist size. However, we mainly use numerical calculation
of Fourier transform, as in such way we are not limited to simple
wave profiles, but can numerically input any field profiles. In general,
it is sufficient to use only $50$ plane waves in the integration
of Eq. \ref{eq:tmm-pw-exp} (trapezoidal integration) in order to
represent common beam profiles reasonably well.

\subsubsection{Power flow of s-polarized wave\label{subsec:tmm-for-waves-pwr-s}}

In the case of s-polarized single plane wave the power flow through
the rectangle with dimensions $L_{x}\times L_{y}$ is easily calculated
from the electrical field amplitude 

\begin{equation}
P_{s}=\frac{1}{2\omega\mu_{0}}\Re\left[k_{z}\right]\left|E_{0y}\right|^{2}L_{x}L_{y}.
\end{equation}

\noindent However, such simple relationship is not applicable for
waves consisting of many plane waves because the calculation of Poynting
vector is a nonlinear operation. In this section, we derive formulas
for calculation of power of s-polarized wave with arbitrary profile,
the results of p-polarized wave are in Sec. \ref{subsec:tmm for waves-pwr-p}.

First, to calculate the power through a rectangle in xy-plane, we
only need to calculate the z-component of Poynting vector. Taking
account that we are currently limited to s-polarization the Poynting
vector is

\begin{equation}
\left\langle \mathbf{S}\right\rangle _{zs}=\frac{1}{2}\Re\left[\mathbf{E}\times\mathbf{H}^{*}\right]_{z}=-\frac{1}{2}\Re\left[E_{y}H_{x}^{*}\right]\label{eq:tmm for waves-poyting}
\end{equation}

\noindent and the corresponding power through the rectangle ($x_{0}..x_{1}$,
$0..L_{y}$) with dimensions $L_{x}\times L_{y}$ is

\begin{equation}
P_{s}=-\frac{L_{y}}{2}\Re\int_{x_{0}}^{x_{1}}dx\,E_{y}H_{x}^{*}.
\end{equation}

Analogous to Eq. (\ref{eq:tmm-pw-exp}) the electric fields can be
expressed as

\begin{equation}
E_{y}=\int_{-k}^{k}dk_{x}\,E_{0y}\left(k_{x}\right)e^{i\left(k_{x}x\pm k_{z}z\right)},
\end{equation}

\noindent where $E_{0y}\left(k_{x}\right)$ is the amplitude of plane
wave calculated by TMM. The magnetic field follows from 
\begin{equation}
\mathbf{H}=\frac{1}{\omega\mu_{0}}\left(\mathbf{k}\times\mathbf{E}\right)
\end{equation}
and the complex conjugate of the x-component is expressed as

\begin{equation}
H_{x}^{*}=\frac{-1}{\omega\mu_{0}}\int_{-k}^{k}dk_{x}\,k_{z}E_{0y}^{*}\left(k_{x}\right)e^{-i\left(k_{x}x\pm k_{z}z\right)}.
\end{equation}

\noindent Thus, the power of the beam is represented by the triple
integral one over the x-coordinate and two over x-component of the
wave-vector. Numerical calculation of such integral requires a fair
amount of computational power, especially the integration over x-axis,
as the electrical and magnetic fields are highly oscillating functions
over space. Fortunately, after rearrangement of integrals it is possible
to analytically calculate the integral over x-coordinate. The expression
for the power of the beam then becomes

\begin{eqnarray}
P_{s} & = & \frac{L_{y}}{2\omega\mu_{0}}\Re\left[I_{ks}\right]\label{eq:tmm for waves-s-pol-pwr}\\
I_{ks} & = & \iint_{-k}^{k}dk_{x}dk_{x}'\,k_{z}'\hat{E}_{0y}^{*}\left(k_{x}'\right)\hat{E}_{0y}\left(k_{x}\right)F_{x}F_{z},\nonumber 
\end{eqnarray}

\noindent where $F_{x}$ and $F_{z}$ describe the interference between
the plane waves in x- and z-direction, respectively. Those coefficients
are defined by

\begin{eqnarray}
F_{x}\left(\Delta k\right) & = & \int_{x_{0}}^{x_{1}}dx\,e^{i\Delta kx}=\frac{-i}{\Delta k}\left.e^{i\Delta kx}\right|_{x_{0}}^{x_{1}}\label{eq:tmm for waves-Fx}\\
F_{z}\left(k_{z},k_{z}'\right) & = & e^{i\left(k_{z}-k_{z}'\right)z},\label{eq:tmm for waves-Fz}
\end{eqnarray}

\noindent where $\Delta k=k_{x}-k_{x}'$. Int the limit $\Delta k\rightarrow0$,
the Eq. (\ref{eq:tmm for waves-Fx}) becomes
\begin{equation}
\lim_{\Delta k\rightarrow0}F_{x}\left(\Delta k\right)=x_{1}-x_{0}=L_{x}.
\end{equation}

\noindent Calculation of double integral in Eq. (\ref{eq:tmm for waves-s-pol-pwr})
is readily done by numerical methods. The calculation of powers is
included to the code of NLTMM and is available at \href{https://github.com/ardiloot/NonlinearTMM}{github.com/ardiloot/NonlinearTMM}. 

\subsubsection{Power flow of p-polarized wave\label{subsec:tmm for waves-pwr-p}}

In case of p-polarization, the derivation stays the same, however
now the main field component is given by $H_{y}.$ The power of the
wave is given by

\begin{eqnarray}
P_{p} & = & \frac{L_{y}}{2\omega\varepsilon_{0}}\Re\left[\frac{I_{kp}}{\varepsilon}\right]\\
I_{kp} & = & \iint_{-k}^{k}dk_{x}dk_{x}'\,\hat{H}_{0y}\left(k_{x}\right)k_{z}\hat{H}_{0y}^{*}\left(k_{x}'\right)F_{x}F_{z},\nonumber 
\end{eqnarray}

\noindent where $F_{x}$ and $F_{z}$ are given by Eqs. (\ref{eq:tmm for waves-Fx})
and (\ref{eq:tmm for waves-Fz}).

\section{Nonlinear transfer-matrix method\label{subsec:NLTMM}}

Standard TMM is used by NLTMM to calculate the electrical fields of
the input beams (non-depleted pump-wave approximation), so the functionality
derived in Sec. \ref{sec:Standard-TMM} is also essential for NLTMM
for realistic beams. However, the calculation of the generated beam
in NLTMM significantly differs from the calculation of the input beams
and is reviewed here.

\subsection{Review}

NLTMM directly solves the Maxwell equations in any nonlinear layered
structure (see Fig. \ref{fig:nltmm-structure}) in the limit of non-depleted
pump wave approximation. Our focus is devoted to the second-order
nonlinear processes (e.g second-harmonic, sum-frequency, difference-frequency
generation), but in general, the method could be extended to the higher-order
nonlinear processes. The main equation for electrical field in nonlinear
homogeneous isotropic medium directly follows from Maxwell equations
and is given by

\begin{equation}
\nabla^{2}\mathbf{E}+\frac{\omega^{2}}{c^{2}}\varepsilon\left(\omega\right)\mathbf{E}=-\frac{\omega^{2}}{\varepsilon_{0}c^{2}}\mathbf{P}_{NL}+\nabla\left(\nabla\cdot\mathbf{E}\right),\label{eq:nltmm-wave equation}
\end{equation}

\noindent where $\mathbf{P}_{NL}$ denotes the nonlinear polarization,
which is the driving term for the generation of nonlinear wave \cite{Boyd2008}.
Symbol $\nabla$ is Nabla-operator, $c$ is the absolute speed of
light and $\varepsilon\left(\omega\right)=n\left(\omega\right)^{2}$
is the relative permittivity of the medium. Eq. \ref{eq:nltmm-wave equation}
describes the fields in a homogeneous layer (layer index omitted,
see Fig. \ref{fig:nltmm-structure}), between the different layers
the fields must connected to match the continuity conditions \cite{Loot2017,Novotny2006}.
For comprehensive review of NLTMM see Ref. \cite{Loot2017a}.

\subsection{Absorption\label{subsec:nltmm-absorption}}

In the case of standard TMM the calculation of absorption could be
easily done through the powers of incident, reflected and transmitted
beam. The calculation of absorption of the generated beam is not possible
in a similar manner because the ``incident/source'' power in not
known. In order to calculate the absorption of the generated wave
the NLTMM was extended. As the nonlinear mediums in this study are
always non-absorbing, it is sufficient to calculate the absorption
of generated fields in linear absorbing layers. The absorbed energy
in rectangular box ($0..L_{x},0..L_{y},z_{0}..z_{1}$) is given by
(coordinates defined in Fig. \ref{fig:nltmm-structure}, layer index
$i$ omitted) 
\begin{equation}
A=\frac{1}{2}\varepsilon_{0}Im\left[\varepsilon\right]\omega L_{x}L_{y}\int_{z_{0}}^{z_{1}}\mathbf{E}\cdot\mathbf{E}^{*}dz,\label{eq:absorbed power}
\end{equation}

\noindent where $\varepsilon_{0}$ is the permittivity of the vacuum
(layer index is omitted) \cite{Novotny2006}. The integral in Eq.
\ref{eq:absorbed power} could be easily analytically solved if the
expression (Eq. \ref{eq:frw-bkw}) for the electrical field vector
is substituted into Eq. \ref{eq:absorbed power} and taking into account
that $\exp\left(i\mathbf{k}^{\pm}\cdot\mathbf{r}\right)=\exp\left(ik_{x}x\pm ik_{z}z\right)$
in our geometry (see Fig. \ref{fig:nltmm-structure}) we arrive to

\begin{eqnarray}
\int_{z_{0}}^{z_{1}}\mathbf{E}\cdot\mathbf{E}^{*}dz & = & \mathbf{-A}^{+}\mathbf{A}^{+*}\cdot\left.i\,\frac{e^{i\left(k_{z}-k_{z}^{*}\right)z}}{k_{z}-k_{z}^{*}}\right|_{z_{0}}^{z_{1}}-\nonumber \\
 &  & -\mathbf{A}^{+}\mathbf{A}^{-*}\cdot\left.i\,\frac{e^{i\left(k_{z}+k_{z}^{*}\right)z}}{k_{z}+k_{z}^{*}}\right|_{z_{0}}^{z_{1}}+\nonumber \\
 &  & +\mathbf{A}^{-}\mathbf{A}^{+*}\cdot\left.i\,\frac{e^{-i\left(k_{z}+k_{z}^{*}\right)z}}{k_{z}+k_{z}^{*}}\right|_{z0}^{z1}+\nonumber \\
 &  & +\mathbf{A}^{-}\mathbf{A}^{-*}\cdot\left.i\,\frac{e^{-i\left(k_{z}-k_{z}^{*}\right)z}}{k_{z}-k_{z}^{*}}\right|_{z0}^{z1}.\label{eq:abs-integralls}
\end{eqnarray}

The possibility to calculate the absorption of the generated nonlinear
wave allows to calculate the out-coupling efficiency, to characterize
the ratio between the power of absorbed (lost) and out-coupled (detected)
light.

\subsection{Realistic beams\label{subsec:NLTMM for waves}}

The calculation of the fields and powers of the generated beam of
second-order nonlinear process is similar to standard TMM. However,
the single integral in Eq. \ref{eq:tmm-pw-exp} is replaced by a double
integral over both input waves $k_{x1}$ and $k_{x2}.$ The plane
wave amplitudes $\mathbf{\check{E}}_{w}\left(k_{x1}',k_{x2}'\right)$
could be readily calculated by NLTMM (Sec. \ref{subsec:NLTMM}).

The calculation of the powers of the generated beam is even simpler
as the formulas described in Sec. \ref{subsec:tmm-for-waves-pwr-s}
and \ref{subsec:tmm for waves-pwr-p} depend only on $\hat{E}_{0y}\left(k_{x}\right)$
and $\hat{H}_{0y}\left(k_{x}\right).$ Those dependencies is readily
calculable by NLTMM. Otherwise, the formulas to calculate the power
of s- and p-polarized beam with arbitrary profile stays exactly the
same.

\section{Discussion and conclusions}

This paper focuses on extending the widely used transfer-matrix method,
used for modeling of layered structures, to beams with realistic dimensions.
The formulations are described in the detail and the code is freely
available at \href{https://github.com/ardiloot/NonlinearTMM}{github.com/ardiloot/NonlinearTMM}.
In addition to the extension of the standard TMM also nonlinear TMM
(developed in Ref. \cite{Loot2017}) is extended to incorporate the
excitation with beams with any profile.

\begin{acknowledgement}

The research was supported by the Estonian research project IUT2-27.
\end{acknowledgement} 

\bibliographystyle{IEEEtran}
\bibliography{references}

\end{document}